\documentclass[runningheads,a4paper]{llncs}
\usepackage{amssymb}
\usepackage{graphicx}
\usepackage{array}
\usepackage{url}
\urldef{\mailsa}\path|{edriosp}@stanford.edu|    
\newcommand{\keywords}[1]{\par\addvspace\baselineskip
\noindent\keywordname\enspace\ignorespaces#1}

\begin{document}
\mainmatter
\title{Spectral Decomposition in Deep Networks for Segmentation of Dynamic Medical Images}
\author{Edgar A. Rios Piedra \and Morteza Mardani \and Frank Ong \and Ukash Nakarmi \and \\ Joseph Y. Cheng \and Shreyas Vasanawala}
\authorrunning{E. Rios Piedra et al.}
\titlerunning{Spatial and spectral decomposition for segmentation of medical images}
\institute{Department of Radiology, Stanford University, CA, United States\\ Department of Electrical Engineering, Stanford University, CA, United States}
\toctitle{}
\tocauthor{}
\maketitle
\begin{abstract}

Dynamic contrast-enhanced magnetic resonance imaging (DCE-MRI) is a widely used multi-phase technique routinely used in clinical practice. DCE and similar datasets of dynamic medical data tend to contain redundant information on the spatial and temporal components that may not be relevant for detection of the object of interest and result in unnecessarily complex computer models with long training times that may also under-perform at test time due to the abundance of noisy heterogeneous data. This work attempts to increase the training efficacy and performance of deep networks by determining redundant information in the spatial and spectral components and show that the performance of segmentation accuracy can be maintained and potentially improved. Reported experiments include the evaluation of training/testing efficacy on a heterogeneous dataset composed of abdominal images of pediatric DCE patients, showing that drastic data reduction (higher than 80\%) can preserve the dynamic information and performance of the segmentation model, while effectively suppressing noise and unwanted portion of the images.

\end{abstract}

\section{Introduction}

Dynamic contrast-enhanced (DCE) MRI is a multi-phase imaging technique that generates time-series images of the body and is widely used in the clinic. These images usually undergo manual segmentation of the object of interest (e.g., organ, tumor, other) to later extract clinically meaningful biological information that is used for diagnosis (e.g., functional status, anatomical characteristics, other bio-markers). In lieu of the manual segmentation, deep neural networks have been proved to outperform previous methods \cite{DL1,DL2} by significant margins, but there are challenges that still need to be solved. \\

Some of these challenges include the complexity of the data that goes into the model, general to the domain of medical images and especially applicable to dynamic information, much of this large amount of data that is acquired can be redundant on both the spatial and temporal components that may unnecessarily increase the complexity of the model and alternatively exponentially increase the time it takes for the radiologist to manually process the data \cite{Seg1}. \\

Additionally, image heterogeneity poses a special challenge in medical image given the variety of sources from which it can originate, including physical artifacts (ringing, chemical shift, ghosting), motion blur (respiratory, cardiac, patient movement), presence of pathological cases, resolution and contrast between tissues, among others \cite{Htr1,Htr2}. Image artifacts and heterogeneous images are inherently more likely to be observed on time-series dynamic data such as DCE-MRI generates large 4D datasets with multiple observations of the area of interest. \\

For this work, we hypothesize that the large dynamic datasets can be more efficiently used by doing a series of spatial and spectral processing to reduce the heterogeneity observed in temporal domain (across phases) and the spatial domain (noise, movement, artifacts). More specifically, the main spectral components together with localized segmentation can lead to a smaller, noise-reduced and time-invariant input for the following deep architecture (U-Net, V-Net, CNNs, etc.) \\

Pediatric kidney DCE-MRI represents an appropriate domain to test this hypothesis given that the proportion of acquired versus useful data is large \cite{DCE}. These kind of data-sets are more likely to be affected by movement over time (infants tend to move more than adults) and additionally, there is a higher degree of heterogeneity that is a result of the presence of some abnormality (e.g., polycystic kidney disease, genetic variations, cysts, tumors, etc.) \cite{Htr1,Htr3}. \\

Reported experiments include the evaluation of training/testing efficacy on a heterogeneous DCE-MRI dataset composed of abdominal (kidney) images of pathologic pediatric patients. Results include the comparison between the performance of a modified U-Net architecture under different scenarios including the reference performance when the full DCE input is used for segmentation, and where a low-rank spectral decomposition at different image spatial scales are utilized. The main contribution of this paper is that were able to show that this methodology can be applied to dynamic medical data-sets and effectively suppress the noise and irrelevant data, while achieving a segmentation performance comparable to that obtained when using the full data.

\subsection{Input data and system architecture} 

\subsubsection{DCE-MRI dataset:}

During a DCE scan, multiple 3D MRI scans (phases) are acquired after the intravenous injection of a contrast agent (e.g., Gadolinium) with the goal of observing its flow throughout the kidney to detect regions of abnormality (e.g., segmentation of the renal compartments) \cite{DCE1,DCE2}. During each phase, the flow of the contrast agent is captured as the relaxation characteristics of nearby tissues change over time and are observed as hyper-intense structures proportionally to the amount of present contrast agent (perfusion through blood flow). Figure 1 (upper rows) shows the overall DCE process. \\

The utilized data-set was collected with IRB approval and consists of 40 high-resolution multi-contrast pediatric DCE cases from which 25 were used for training, 5 for validation and 10 for test. These cases included some degree of abnormality (e.g., hydronephrosis, polycystic kidney disease, congenital anomalies) that introduce heterogeneity to our dataset. Manually delineated regions of interest (ROI) were generated for both kidneys by an expert technologist, with subsequent radiologist editing, to train and assess system performance. \\

Imaging was performed using a multi-phase 3D modified SPGR sequence with motion navigation, intermittent spectrally selective fat-inversion pulses, and VDRad sampling patterns were used during the contrast injection. Minimum echo time (TE) 1.2–1.6 ms, repetition time (TR) 3.0–3.7 ms, flip angle 15 degrees bandwidth (BW) 100 kHz, slice thickness 0.9–1.2 mm, FOV 20–44 cm, spatial resolution 0.8 x.8–1.4 x 1.4 \(mm^2\), and a total acceleration factor of 7.8–8.0. A total of 50 phases of 100 images each were acquired for each case.

\begin{figure}
\includegraphics[height=8.2cm]{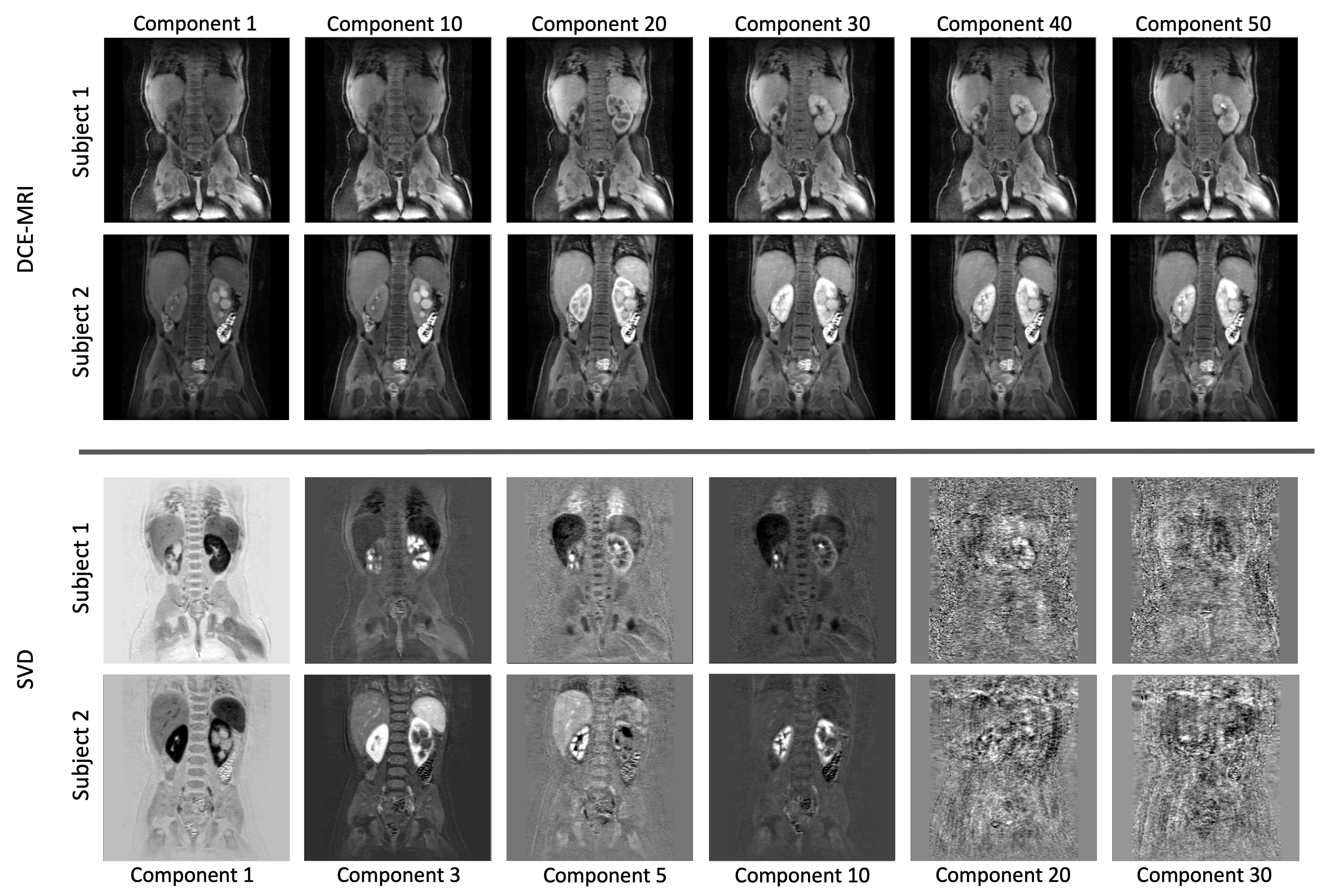}
\caption{Comparison of DCE and spectral decomposition. The upper rows show some of the phases that compose the 4D DCE-MRI. It can be observed how the contrast agent flows from the vascular system to perfuse the kidneys (cortex then the inner medulla) and later go on to the collector system. The lower rows show the reconstruction of the images once the main components have been calculated using SVD. Different areas of hyperintensity can be seen in the first sets of components, representing the dynamic enhancement with reduced heterogeneity. Background noise is captured in the latter components.}
\label{fig:DCESVD}
\end{figure}

\subsubsection{Network architecture:}

The utilized architecture is multi-channel U-NET based on \cite{UNET}, as it has been shown to have a robust performance for the production of accurate results in volumetric medical imaging scenarios \cite{Seg1,Seg2,Seg3}. The model consists of three 3x3 convolution layers on the contracting path and a ReLU and max pooling layers (stride 2) at the end of each block. The upstream network presents a similar configuration with 2x2 up-convolution step and 3x3 convolution layers. The final layer at the end of the up-stream side is a 1x1 fully connected layer to produce pixel-wise scores for the ROIs of size \(x_i,y_i\) for an N number of slices that matches the input image volumes. An illustration of the network and its different inputs is shown in Figure 2. \\

A cross-entropy loss function and the Adam optimizer using Keras with Tensorflow on an NVIDIA GTX TITAN GPU with images of size 256x256 pixels using patches of size 48, 24, 12 and 6 on both down- and up-streams for 100 epochs (early-stop enabled).

\section{Methods}

This section describes the process to perform a segmentation at multiple image scales around the kidney region, spectral decomposition of the DCE-MRI into its main components, the different combinations tested for each scenario as a different set of \( n \) main components are used to train the system and the comparison with the reference segmentation performance utilizing the full DCE data.

\subsection{Low-rank spectral decomposition}

To decompose the input dynamic dataset we employed the well-known singular value decomposition (SVD) method \cite{SVD}. This method can be used to separate the signal into its main components with respect to their contribution to the overall variation encoded in the image. In the case of dynamic DCE-MRI the input matrix is reconstructed into a 2D input of shape \(  P*(x*y*z)  \), where P is the number of phases (50 in this case) and x, y, and z represent the 3D dimensions of each phase. Then the singular value decomposition M is given by 

\begin{equation} M = U*\Sigma*V^T \end{equation}

where \(  U  \) is has the left singular vectors, \(  \Sigma  \) is rank 1 matrix with singular values (diagonal matrix) and \( V^T  \) contains the right singular vectors. \\

Afterward, the information contained in \(  U  \) ordered according to the relative importance of each component encoded in \(  \Sigma  \) can be reshaped into SVD input images as its coefficients determine how much information from the input is retained in each component. In summary, the first (low-rank) components contain the most relevant information observed across the 50 phases, leaving noise (random information) and artifacts in the last components as these provide the least information. Consequently, the SVD output is time-invariant as the contrast enhancement information is aggregated the main components. Figure 1 (lower rows) shows the output images from this step, which are used as input to the deep-learning model.

\subsection{Local spatial partitioning}

Spatial data reduction can further help to limit the regions where the feature learning occurs and allow for more specific texture patterns to be detected. For this purpose, a localized-segmentation approach that segments the input images at multiple scales was utilized. The overall idea is to center the learning process on the fragments that contain the region of interest and test the performance under different image partitions or sizes during training. This approach has been determined in MRI and other image domains to be useful in cases where local features can provide better results than the patterns observed at the global scale  \cite{MS1,MS2}. \\

Three different iterations of the network were created on which increasing locality regions were utilized, each of them was centered around the kidney region (structure of interest) going from regions that included only the internal kidney components (e.g., medulla, cortex), a medium one that contained some of the peripheries outside the kidney cortex and a bigger one that contained a wider region of the thoracic region \cite{MS3}. Figure 2 shows the input architecture as well as the multi-scale and SVD inputs to the network. For each case, a kidney region of interest (ROI) was found and then compared to the manual gold standard to obtain the segmentation accuracy according to the Dice coefficient \cite{DICE}.

\begin{equation} D = \frac {(2 * TP )} { ( 2*TP + FP + FN)}    =   \frac {2 * |A \cap B| } {(|A | + | B |  )} \end{equation}

where D is the Dice coefficient, A is the automatically generated image and B is the manually generated gold standard.

\begin{figure}
\centering
\includegraphics[width=12.2cm]{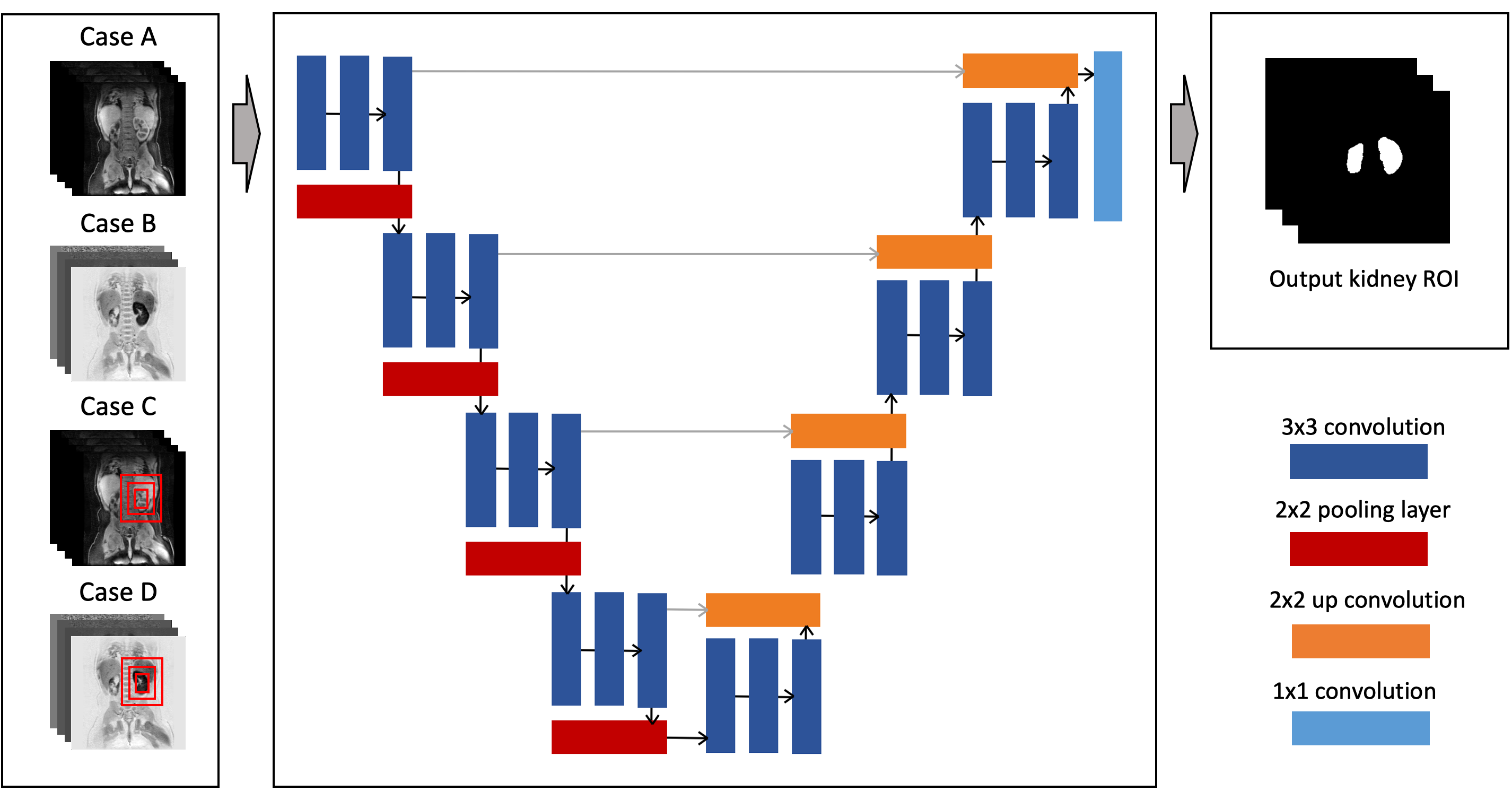}
\caption{Caption: Overall diagram of the system architecture and the different inputs utilized for the experiments. A modified U-Net based on \cite{UNET} was utilized to obtain the ROI. Case A represents the inputs of the original DCE-MRI images. Case B represents the different components generated by the SVD. Case C and D show the case where different sub-scales of the images around the region of interest.}
\label{fig:UNET}
\end{figure}

\section{Results}

The overall process included the training and evaluation of the individual and combined performance of the spectral and spatial decomposition operations with the deep neural network as the number of main input components was increased in each evaluation. We tested all first ten main components and then increased in steps of five until they were all included (fifty different phases for these experiments). A five-fold cross-validation was also performed to avoid model overfit. Finally, Each of the experiments was compared to the performance observed when training and segmenting the images using the all phases of the original DCE-MRI (shown in blue in Figure 4 and the second column in Table 1). \\

Figure 4 contains the distribution of results observed for the spectral decomposition experiments, through the training and testing sets it was observed that the best performance was obtained when using an interval that included the first 10 components. We also observed that in many cases the performance observed was comparable or higher than the segmentation performance using all DCE phases but the performance obtained across all test subjects was about 10\% lower than the reference performance (0.6342 vs 0.7333) but it is worth noting that this was achieved using less than 20\% of the available (8 out of 50 components). Lastly, results after the 25\textsuperscript{th} component provided poorer performance (in practice, adding noise to the training) and were omitted for better data visualization. \\

Lastly, table 1 shows the summarized results using localized segmentation, spectral decomposition, and a combination of both. Showing that at training stage, using the first main components (SVD column) can achieve a better performance than the reference DCE. Comparatively, the localized segmentation did not manage to out-perform the reference in any scenario but managed to increase its performance when applied to the spectral data (MS-SVD column), specifically in the case where the intermediate scale was used (covering kidney medulla, cortex, and immediate kidney periphery). \\

\begin{figure}
\centering
\includegraphics[height=2.5cm]{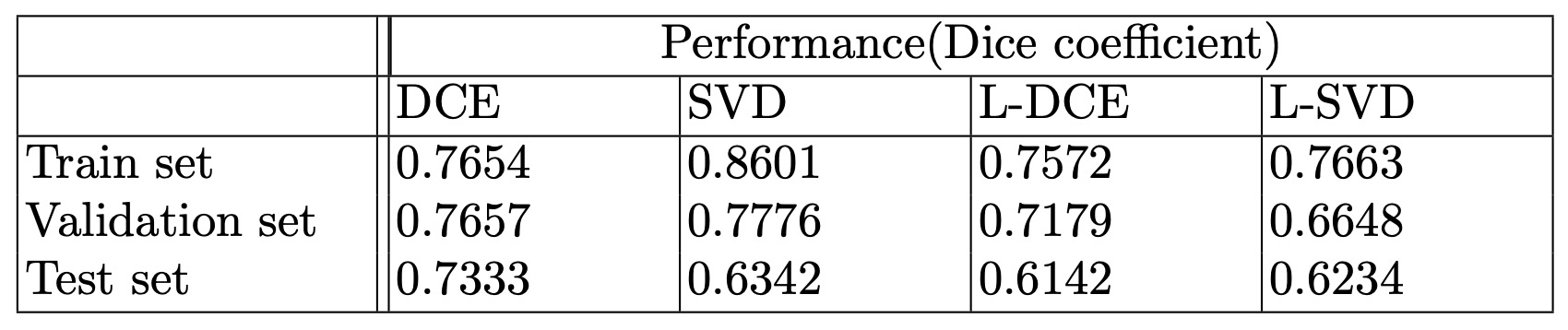}
\caption{Table 1. Summary of results obtained on the different experiments for train, validation and test sets. The DCE and L-DCE column shows the performance when all 50 DCE-MRI phases are utilized for training. SVD and L-SVD show the best performance obtained when using the first eight input components, as commented in the discussion section. L-DCE = Localized DCE, L-SVD = Localized SVD.}
\label{fig:Table}
\end{figure}

\begin{figure}
\centering
\includegraphics[height=13cm]{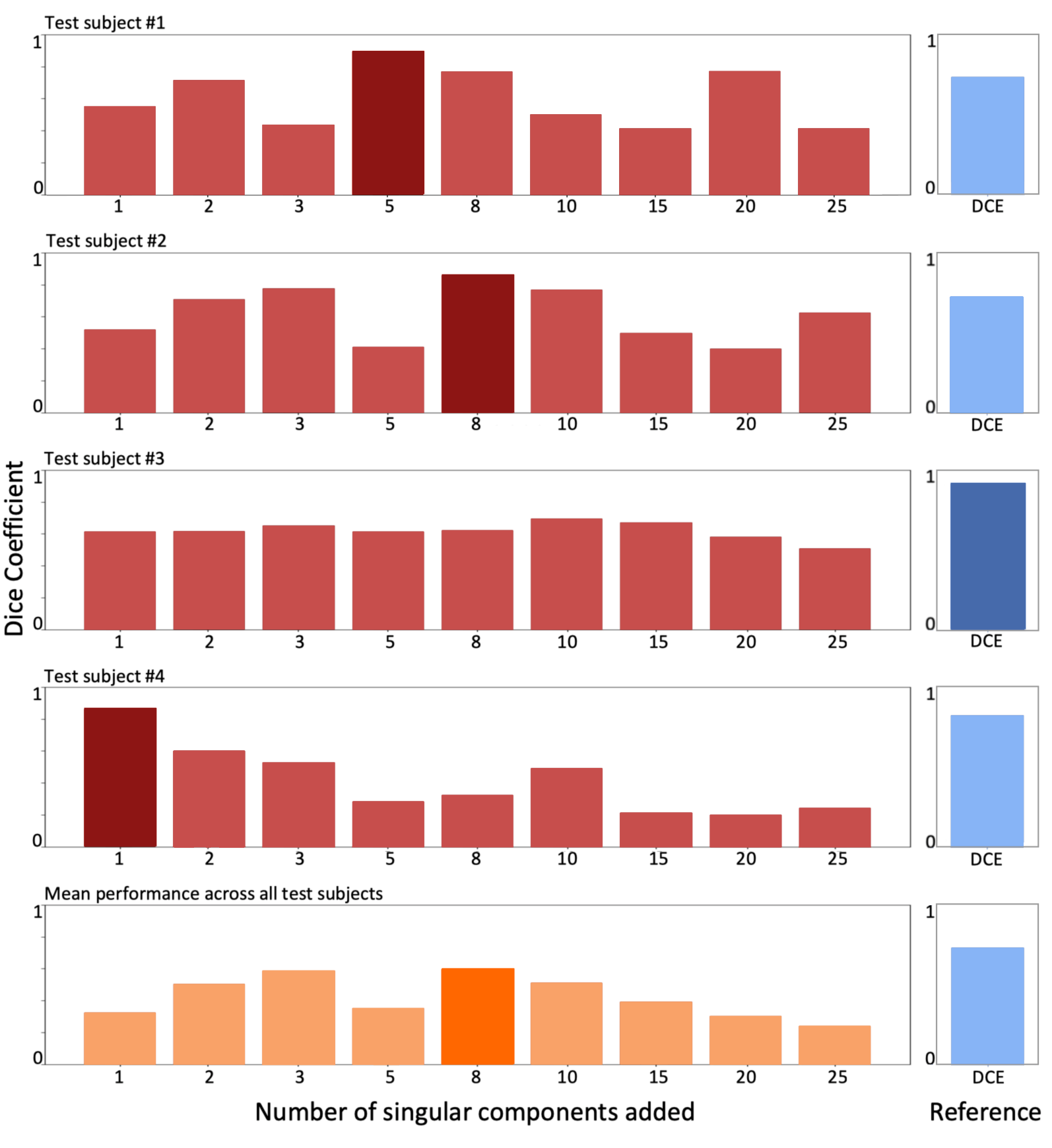}
\caption{Performance of a sample test subjects and average results obtained for all test cases as different singular values are added into the network. Performance is measured as the Dice coefficient between the predicted mask against the expert manual segmentation. The best performing result for each case is shown with stronger color.}
\label{fig:Barchart}
\end{figure}

\section{Discussion and conclusion}

In this project, we attempted to use spatial data reduction and spectral decomposition to constrain the amount of information used from heterogeneous data-sets and evaluate the effect in segmentation accuracy. A multiple scaling approach was utilized to evaluate the spatial utility of different localities of the input data-set. Additionally, we utilized low-rank singular value decomposition to evaluate the spectral information that is relevant for the input kidney data (number of singular values to use) to de-noise, reduce the data input size, and make the input time invariant. These experiments were performed using a modified U-Net architecture, results (Table 1) were observed to achieve a segmentation performance similar to the reference DCE-MRI performance using the full data. The main contribution of this paper is that were able to show that spatial and spectral decomposition can be applied to dynamic medical data-sets to effectively suppress noise and irrelevant data, achieving a segmentation performance similar to the reference full-data model (less than 15\% difference) using only a fraction (8 main components, 16\%) of the original data size, which can be of  extreme utility when working with big sets of dynamic medical data and other applications such as functional MRI, diffusion scans and other non-medical applications. \\

\end{document}